# Privileged Data within Digital Evidence

Dominique Fleurbaaij
Dutch Police,
The Netherlands
dominique.fleurbaaij@ucdconnect.ie

Mark Scanlon
School of Computer Science,
University College Dublin,
Ireland.
mark.scanlon@ucd.ie

Nhien-An Le-Khac
School of Computer Science,
University College Dublin,
Ireland.
an.lekhac@ucd.ie

*Abstract*—**In recent years the use of digital communication has increased. This also increased the chance to find privileged data in the digital evidence. Privileged data is protected by law from viewing by anyone other than the client. It is up to the digital investigator to handle this privileged data properly without being able to view the contents. Procedures on handling this information are available, but do not provide any practical information nor is it known how effective filtering is. The objective of this paper is to describe the handling of privileged data in the current digital forensic tools and the creation of a script within the digital forensic tool Nuix. The script automates the handling of privileged data to minimize the exposure of the contents to the digital investigator. The script also utilizes technology within Nuix that extends the automated search of identical privileged document to relate files based on their contents. A comparison of the 'traditional' ways of filtering within the digital forensic tools and the script written in Nuix showed that digital forensic tools are still limited when used on privileged data. The script manages to increase the effectiveness as direct result of the use of relations based on file content.**

I. INTRODUCTION

One of the problems in digital investigations is the increasing amount of data required to investigate, as storage capacity of individual devices and the number of devices seized per household increase [1, 2]. This is due to the current trend of the digitalization of society, where computing and digital communication is everywhere [3], and storage is rapidly becoming cheaper [4].

The transformation from paper to digital is also setting in on practices where the digital contents are restricted by government laws. Privileged data, such as an e-mail conversation between a lawyer and its client is not to be viewed by third parties. But law enforcement may encounter such information stored on a suspect's seized computer or the e-mail database export from a company.

A police department in the Netherlands has three dedicated teams appointed with the task of investigating large scale financial and economical fraud cases. These investigations often take years to complete due to the complexity that is introduced. This is due to the large amount of information that is required to analyze in order to find the criminal activity, which is often concealed.

Within these investigations, a team is involved which consists of tactical investigators, specialized financial investigators and a digital investigator. The task of the digital investigator is to analyze the physical and digital evidence for information related to the committed crime.

It is stated that the origin of money can be traced, thus it should be easy to find the original source of money that is derived from criminal activities. Unfortunately, criminal activities are more organized, more sophisticated with the help of professional experts. The money can flow through an international network of companies (or corporations) established in countries where there is no law that criminalizes the original crime or where there is no collaboration between the involved countries, which makes the paper trail hard to follow.

In many cases our suspects were helped with the establishment of their companies by advisors such as lawyers, notaries, accountants or tax consultants. They provided these complex services as this is their profession and do not have to be part of the criminal activity. However, their services could be used in order to create the money laundering construction.

Within the Netherlands and most other parts of the world [4], these professions have a specialized status regarding the information they work with from their profession: it's privileged. All forms of communication that are created, send to or from a person that has the right of privileged communication may not be viewed by anyone else than the persons for whom the communication was intended. This includes law enforcement, except under very exceptional circumstances.

During any investigation, often tens of terabytes data are seized in total millions of files. The chance that one of these files is privileged is always present.

Currently there are procedures written by different parties describing how to handle privileged information when it is encountered in the physical and digital evidence. It is very likely that these procedures are written based on the requirements by law, but do they match the possibilities that are given by the digital forensic tools used to view the digital evidence? Some of these tools have the possibility to block further viewing of files, but they are not all suitable for cases where it is required to block privileged files. The implementation of such functionality is left to the developer of the forensic software and has often a limited usability as discussed further in this research.

After encountering a privileged file, a search is made within the forensic tool based on very limited amount of information given by a privileged officer (a different person then the digital investigator) may see of the document. This is often no more than a header of a letter or the email address from or to whom the email is send. The results of these searches are tested again for privileged information and, where applicable, the files from the result are blocked for further viewing within the case. This approach is very limited and imprecise due to the lack of information to create a filter that consists of quality information to find the relevant privileged data.

This approach to filtering the evidence requires a lot of time spend on searching, filtering and blocking the files from viewing. The process is often done on very large datasets which require a large amount of time before the results of each step is presented within the forensic software. In this paper, we aim to answer the following research questions:

- What are the possibilities of common digital forensic software tools regarding the blocking of privileged information?
- Can relevant privileged data be filtered based on content, other than the limited amount of information available due to the superficially analysis restriction that the law enforces when reviewing privileged data?
- Can the process of searching, filtering and blocking items be automated in order to save time and be more efficient?

Some digital forensic software tools such as FTK and Nuix already have been provided with the possibility of blocking the information from further viewing. Nuix also provides extended possibilities to relate files together based not only on their cryptographic signature but also by their content. The contribution of this paper is a Nuix script that can:

- Find duplicate, related and derived files based on a starting position, such as a selected item or an e-mail address.
- Block these files from further viewing.
- Export the privileged files, including a generated summary file list, so that the privileged files can be reviewed if their status is truly privileged.

The rest of this paper is organised as follows: Section II shows the background on legal professional privilege and the removing data from a forensic image. We present our method to identify privileged documents in Section III. We describe our process and script in Section IV. We evaluate and analyze our approach in Section V. Finally, we conclude and discuss on future work in Section VI.

## II. BACKGROUND

### A. Legal professional privilege

Any person should be able, without constraints, to consult a lawyer, whose profession entails the giving of independent legal advice to all those in need of it. Legal Professional Privilege (LPP) is a general principle of law that describes the protection of all communication between a professional adviser and their client and the right of nondisclosure. The privilege can be waived by the client or the lawyer [6]. The recognition of privileged information exists since 1982 [5]. In addition to this privilege, there is also the derived privilege for individuals who are in the employ of, or work for, professionals entitled to privilege, and who thus, by virtue of their office, handle information that is privileged. Examples are a secretary in the employ of a civil law notary or an expert provided with information by a solicitor who is seeking a professional opinion. When examining seized digital information, a digital investigator may come across documents drawn up by solicitors or civil law notaries, who are all bound to secrecy and have the right of nondisclosure. The contents of the documents in question may not be passed on to the judicial authorities, other than what necessary to determine whether (parts of) these documents are privileged [7].

The law mentions letters, writing, and written documents, and in accordance with jurisprudence, this also applies to computer files [8]. The data carrier or forensic image, the container that holds the privileged data, is not privileged [9]. The problem with encountering files in forensic images is that it's difficult to examine the files as separate entities as they are bound by the image file. Therefore, it is important to examine data carriers in such a way that the complainant's right of nondisclosure is maintained [10]. When it is reasonable to suspect that the digital evidence contains a lot of privileged data, there is the option to clean up the seized information before it is passed on to the investigating team.

Besides, there is a problem with the identification of privileged documents: there is no identification requirement when documents are send outside of the privileged environment. The manual from the Dutch bar association with guidelines for electronic communications and the Internet [11] contains a Dutch translation of the directive regarding the handling e-mail and documents set in 2005 by the CCBE. It contains a chapter for using descriptive metadata [12] in documents in order to maintain the authenticity and to set parameters like the author, subject, the process of which the document is a product and the date. This metadata is however not used to mark a document as identifiable as written by a profession with a privileged status.

### B. Removing data from a forensic image

A lot of effort is undertaken by the digital investigator to protect the digital evidence against invalidation [22]. Invalidated evidence because of a modification of the image could lead to the dismissal of any evidence that resided on the forensic image. Although every change can be documented in order to protect the chain-of-evidence, it should not be common practice to break the forensic integrity that the imaging software provides. Privileged data might be a reason why it would be required to remove files from a forensic image. As described in chapter 2.1, the official procedure for removing privileged data states that, if possible, a copy of the image should be made excluding the privileged files.

When a privileged document is found within the physical evidence, it is taken out of the evidence, placed inside an envelope and transferred to the privileged officer for review. The original physical evidence remains intact and there is no question that the investigative team had any possible way of viewing the contents of the document other than the possible header by the person who found it.

By removing the privileged data from the forensic image in the same principle as in the physical evidence, the possibility of reading the privileged data in a digital forensic tool or in the image itself is removed.

*Possibility of removing data from an image*: A possible way in order to remove data from an image is to load the RAW disk image into a hex editor such as HxD [13] and manually overwrite the sectors on the image where the privileged data resides. The altered image can be saved as a copy.

*Digital forensic tools*: It will be most likely that privileged data is found during the investigation of an image in a digital forensic tool. After the image is cleaned of any privileged data, the image should be used as new source within the forensic tools. It might be possible to redirect forensic software tools to use the altered image. This could lead to some unforeseen problems such as unstable software when a piece of removed information which is no longer available is requested by the forensic software in the altered image. An index database, such as that used by the forensic software FTK [14], will still contain the metadata of an e-mail such as the sender and receiver, subject and dates until it is regenerated. Removing the privileged data from an image and re-indexing the new image every time a new privileged document is found would take an extensive amount of time, while the end result by using a filter in the forensic tools would often suffice to hide the information.

*Forensic format*: Forensic images are often stored in the Expert Witness Format (E01) because of the disk space it can save when the compression is used [15]. The downside of compression is that the contents of the image cannot be viewed directly by a hex editor tool. The contents of the image must first be mounted as a writeable disk image or converted to a RAW disk image.

*Integrity:* Images made from a storage device are created forensically sound and their integrity can be checked by a hash value that was calculated during the imaging process. When the content of the image is altered, the hash value of the image changes. Verifying the previously calculated hash will no longer result in a positive check. Another way of integrity testing must be applied.

In the paper Protecting Digital Legal Professional Privilege (LPP) Data [16] the integrity issue is questioned from data that were removed from an image. The paper states that it is possible to verify the individual files on a logical level within the image. However, it is not possible to verify files on a physical level This physical level may still be of importance, for example when deleted files are recovered from the unallocated space. This information does not exist on a logical level. The paper states 'To overcome the problem, the simplest way is to pre-calculate all the physical sector hash of the original storage device and use that hash set to verify the physical level data at the modified image'. This would solve the problem of being unable to verify the data integrity after removing sectors containing privileged information. The problem of this verification method is that, during the writing of this paper, there were no software tools capable of running such a verification.

*Content containers*: There is a possibility that the privileged files will reside inside some form of container. Examples of these could be ZIP files, database files or a PST file containing e-mail. These files should be handled separately. It is vital to forensically extract the contents of the file and remove only the necessary data in order not to corrupt the container file itself and thus making all the contents unreadable. The extracted information could be provided separately to the altered image. Such a change would require testing in order to see what kind of information would be removed with the extraction.

*C. Nuix*

Nuix presents the Investigator Workstation as the main product for the Digital Forensic investigator: a single application built to process large volumes of data [17]. At the time of testing version 7.0.2. was available. NUIX provides a full list of the supported file types [18]. Support at that time for the required file types is present and images and PDF files can be processed with OCR in order to have their contents indexed. The Investigator Workstation has no multi-user environment and can therefore not block or hide files from other users. Files can be placed in an excluded list, where they will be excluded from searches, results view (a file list view) and document navigator (overview of the evidence files in tree-view). The privileged items moved to the excluded item section can still be seen when the excluded items section is specifically selected by the user.

Nuix implemented technology (Shingles) that enables the comparison of identical documents in a way that the investigator could not: by comparing the contents of privileged data. The results are file relations between files that are not identical, but similar. For example, a Word document can be saved as PDF and contain the exact content, but with a different cryptographic hash identity. This file is very relevant, but there is a high chance it will be missed with the filtration process.

To make use of this technology Nuix requires that evidence is processed with the option Enable near-duplicates enabled. When enabled, Nuix generates sets of shingles for each item it processes. A shingle is a series of phrases, typically five words with an overlap on either end [19].

Nuix's ability to relate files based on shingles gives an automated process the possibility to potentially find more privileged files based on their contents. This gives an advantage over the other tools.

## III. IDENTIFICATION PATTERNS

To identify privileged documents or to be able to chain together emails or documents in our context, it is required to define a set of patterns that can be used to create filters within the forensic software.

*Telephone number*: The Dutch law has declared that since the 1st of September 2011 the following telephone numbers that are used by persons with a privileged profession must be recorded in a central database [23]:

- The extension number of its fixed telephone

- The extension number of the fax machine that is used only by the lawyer, other privileged holders or persons with a derivative of his privilege

- The extension number of the secretary of the lawyer

- The number of a fixed (separate) telephone in the home of the lawyer, provided this terminal is intended solely for business use.

- When a phone number is being tapped and a call is made from or to a privileged phone number, the system that records the call will automatically remove calls made from these numbers.

***E-mail***: In contrast to the telephone number, there is no requirement for a person with a privileged profession to register his e-mail address or domain name officially. There is no possibility to check for privileged e-mail addresses. E-mail communication can be traced by the information present in the header. These headers contain, among other fields, the FROM and TO e-mail addresses of the sender and recipient(s) and might contain e-mail addresses in the CC and BCC fields. This information is interpreted by digital forensic tools and makes them available for searching. Finding relations between privileged files should not be done with the Carbon Copy (CC) and the Blind Carbon Copy fields.

This concludes that we do not require to filter for privileged files when the privileged person is in the CC and the BCC will most likely be too unreliable for identification. The digital investigator should be aware of the following relations:

- In the case of a privileged e-mail it should not be forgotten to block files attached to the original e-mail.

- Filtering on the privileged email address could identify a lot more privileged e-mails.

- After the email address is checked, a domain name check should give away if there was any derivative privileged communication.

- In the case of a file, there should be a search throughout the case in order to find documents that are identical. This could be done based on the hash value of the file, which is often generated during the index process in the digital forensic software

- There is a possibility that the content of a document is stored in another format. For example; the original Word version of a PDF might still be stored on the computer, but will not be found during a hash based search. Ideally the contents of the file should be tested against other documents for overlapping word phrases within the case.

***Keywords:*** To complete the search for privileged communication, a keyword search can be done. This search should include the following:

- Full name: John Doe

- Surname: Doe

- Keywords that fulfil a certain probability: lawyer, notary, confidential, etc.

A keyword based search will only be an advantage if the keywords are known and specific enough. The term lawyer is widely used on modern day computers by software packages that come with legal documents. In order to use keywords with privileged professions, it would require a source, e.g., a database, containing company names, e-mail addresses or telephone numbers of privileged communicators. Due to the typical precision/recall tradeoff in the field of information retrieval, certain broader queries could result in a large non-relevant percentage of discovered documents in the result [20]. A proactive keyword search without very specific data that identifies privileged persons will result in false positives and could potentially remove evidence from the case. In 1985, an evaluation was conducted on the retrieval effectiveness of a full-text document retrieval system [21]. The results where that less than 20 percent of the documents relevant to the particular search were retrieved.

File relations: File relations can be used when a privileged file has been found, either through a search containing specific keywords or by stumbling upon it. It is recommended that the relationship to other files is expanded in order to view the possible chain of communication. A single privileged file might contain enough metadata to find the related files. A relation between different files can be made based on the metadata within a document, such as the FROM email address inside the header of an email or the Author metadata added to a Word document. These properties can be analysed without reading the contents of the file and used to expand the search for related documents. In Figure 1, a fragment is shown on how items could possibly relate to each other.

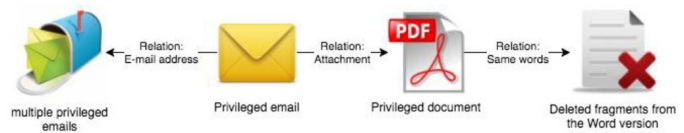

Fig. 1. Public keys associated with the wallet's transactions, which occurred as binary values in memory

## IV. PROPOSED PROCESS

A proof-of-concept script for Nuix was written and outlined in this paper. Its goal is to automate the handling of privileged data. The script automates the process of searching, blocking and exporting privileged data. For a more thorough result, the script uses technology in Nuix to find object relations and compares file contents based on Shingles described in Section II.C.

The script is to be called upon after finding an e-mail or document that contained privileged information. It is written to start from two scenarios:

1. One or more items are found containing privileged information and the investigator is required to filter these from the case. These items are selected within the GUI and the script is started.

2. The investigator has information that a certain e-mail address is used by a privileged user and wants to filter on this entry.

The flowchart in Figure 2 represents a high-level workflow of the script. The script can be separated in three different stages; start-up, processing and export. Each of these steps will be explained in the following sub-sections.

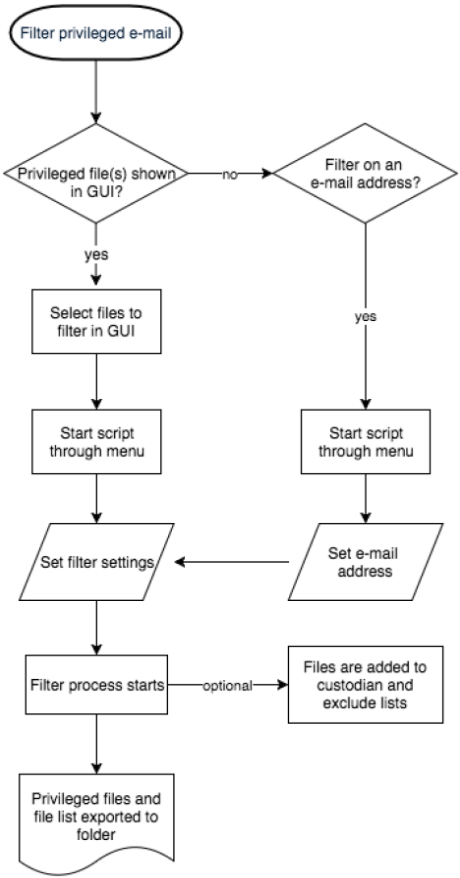

Fig. 2. High-level workflow of the script

*A. Start-up*

Before the script is started, a few settings are required to start. These parameters allow to alter the result and output of the script in order to suite the requirements he or she might have. The script will set default parameters upon each start.

The first step after the script has been started is a check if there are items in the case of the media type application/pdf that do not contain text within the Nuix database. Media Types (MIME types) are two-part identifiers for file formats and are used within Nuix to identify the type of a file.

If unsearchable PDF documents are present in the case, a warning message is shown to the user that it is advised to run Optical Character Recognition over these items. Without this step, it is possible that unsearchable PDF documents are not matched as related, as their content cannot be read.

Next, the user is required to enter data into three windows:

1. The export directory will be used to store all the found privileged items. The script defaults to the user's Desktop folder:

   ~/Desktop/NUIX_PrivilegedItems_Export

2. The required amount of shingles (file similarity) to apply during the search. The default is 0.9, which means that files are only related to each other if 90% of the file content is similar.

3. Should the script create Custodians and Exclude lists for the items that will be found inside the Nuix case. A Nuix custodian is used to group items under a specified name. Items contained in a Nuix exclude list will be suppressed from the results view and the document navigator. This automatically means that they will also not appear during searches. This function is implemented to test-run the script without making any alterations to the Nuix case. The creation of a custodian and exclude list is the only alteration the script can make to the case.

*B. Processing*

If the script is started without the selection of one or more files in the Nuix GUI, it is assumed that the user wants to filter on an e-mail address. A popup window allows the user to enter this information.

The script starts a search based on an e-mail address or selected files. The result of this search is placed inside a queue. For each item in the queue a search is started to find related files based on duplicates, near-duplicates and chained-duplicates. This related file search is based on the MD5 digest of a file and files that are above the resemblance threshold of shingles setting. In the case of e-mail, the related search is only performed on the FROM e-mail address. Individual files that were selected through the Nuix GUI are processed to find related items based on the same methods as use to find duplicates in e-mail, as can be seen in Figure 3.

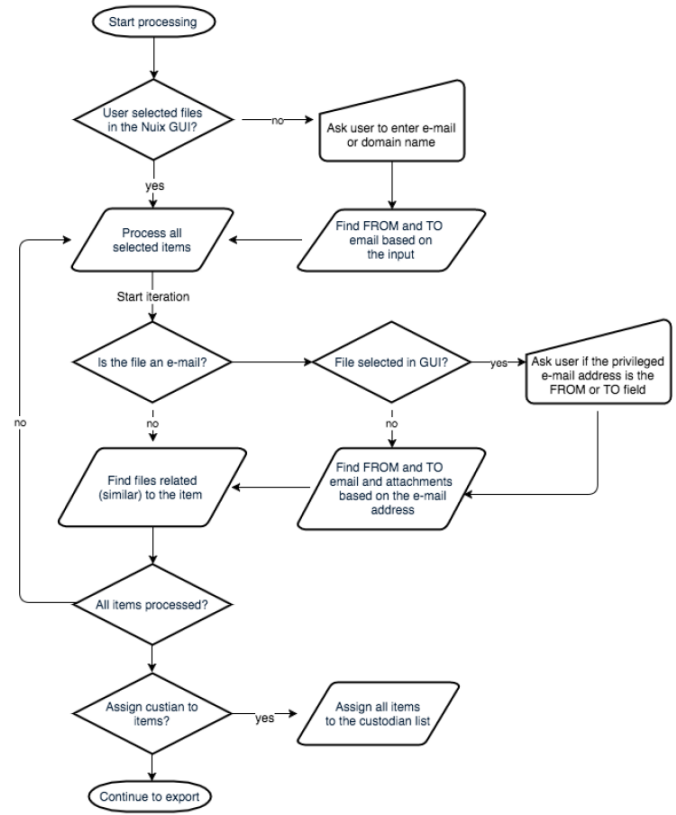

Fig. 3. Script processing workflow

During development tests, it was discovered that evidence could be flagged privileged while this was not the case. This happened when a relation search was performed on the TO e-mail address. For example; when a suspect sends an e-mail with multiple users in the TO field and one of these recipients is our filter e-mail address, all the e-mails to the non-privileged users will be flagged privileged because their content is the same. Therefore, the file related search runs only on the FROM (privileged) e-mail address.

As can be seen in Figure 4, the investigator does not get to see any information such as the subject of an e-mail. Everything is handled outside the view of the investigator. When all items are processed, items are assigned to custodian- and exclude lists, if the user has required this in the start-up configuration.

Fig. 4. A part of the script output when processing an e-mail address

## C. Exporting

The Nuix API contains a function that can convert each item into a PDF representation. This function is applied by the script to generated PDF files that can easily be viewed by a third party without the requirement of any digital forensic software.

The generated PDF files are stored with filenames that are derived from an incremental number and the type of content the original item was, such as e-mail, document, image, etc. This way the investigator does not have any information regarding the content of the file, but a magistrate could know what kind of document is to be expected, as can be seen in Figure 5.

Fig. 5. Exporting every file as PDF with a custom name

In the same folder as the PDF files, a tabbed file list is generated containing the following fields:

- Export Filename - Filename used for the exported file
- NUIX GUID - Used to uniquely identify the exported file within the NUIX case
- Original File Name - Original name of the file as found in the evidence file
- Original File MD5 hash - MD5 digest as calculated by NUIX. This is done over the file in de digital evidence, not on the exported PDF file.
- Original File Path - Original file path as found in the evidence file
- Export status Contains the status of the file export to PDF. A file can appear in the file list, but not in the export directory as PDF. This could happen when a file could not be converted to PDF, such as binary files or container files such as ZIP

The generated PDF files along the file list are written to the export folder, as can be seen in Figure 6. After this the script will end with a message window that informs the user that it is complete and the total number of privileged files that are processed.

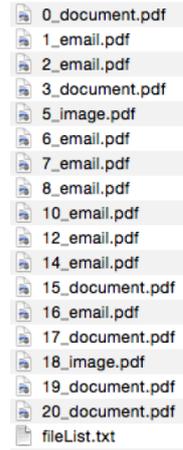

Fig. 6. Exported PDF files together with a file list containing the source of each PDF

## V. EXPERIMENTAL RESULTS

For the purpose of testing, a setup of 3 virtual machines containing simulated privileged communication and files were created. With this setup, three forensic environments are tested; The Nuix script, the Nuix GUI and FTK. The Nuix GUI is the normal interface of the Nuix Investigator Workstation in which the digital investigator works. Autopsy is left out of the comparison as cannot be used to block privileged files. The test case contains a total of 23 uniquely generated items. Out of these 23, 13 unique items are privileged; the e-mails and their attachments from the lawyer and vice versa, together with 5 desktop files which are duplicates from the privileged attachments. The results are outlined in Table 1.

TABLE I. TEST CASES

|  | Total | Privileged |
|---|---|---|
| Total unique items | 23 | 13 |
| Emails | 14 | 7 |
| Attachments | 7 | 4 |
| Desktop files | 5 | 5 |

The total number of unique items is not equal to the sum of e-mails, attachments and desktop files as some files are used multiple times.

*Results FTK:* A filter was created within FTK that contained the lawyer e-mail address on the FROM and TO field. This resulted in the display of 20 e-mail messages. Of these 20 e-mail messages, 6 e-mails where unique. During the analysis of the results it was noted that FTK scored a high number of total items, but a very low number of unique e-mails in comparison to the Nuix GUI. After further research it was found that FTK has hit 6 send e-mails twice in the displayed results. The double entries are removed from the statistics.

FTK was unable to display attachments belonging to the lawyer e-mail address with any of the possible filters. Attempts were made together with the support of AccessData to show the attachments of e-mails (child items), but this was not possible within FTK.

*Results Nuix GUI*: The privileged communication was found by creating a search query that uses the fields from-mail-address and to-mail-address of the lawyer. This resulted in a total of 22 items, of which 10 unique. The result contains 6 privileged e-mails and their corresponding 4 attachments. The result is expected, as the 5 desktop files cannot be related to the e-mail communication.

*Results Nuix Script*: The script is started with a shingle setting of 0.90 to relate only files of which their content is minimally 90% the same. The e-mail address of the lawyer is used as filter criteria. The search uses the same fields as used in the Nuix GUI, but an extra file relation search is run on each of the resulting files.

The result is a total of 23 files are found, of which 13 unique. The results contain 7 e-mails and their 4 corresponding attachments. The script was able to filter out related files such as file1_1.docx, which has 5 words less than the original file1.docx and file1_2.pdf which is the PDF version of file1.docx based on their contents.

Conclusion: In the table below are the statistics of the three test performed to filter on an e-mail address with a privileged status (Table 2).

The Nuix script has found all (100%) unique privileged items in comparison to the 10 items in Nuix GUI (76.9%) and the 6 items by FTK (46.15%).

Comparing the total number of found privileged items, the Nuix GUI finds 68.75% and FTK 43.75% of privileged items when compared to the Nuix script.

TABLE II.  COMPARISON ON THE PRIVILEGES STATUS FROM THREE TESTS

| Privileged items<br>( ) = unique items | Nuix GUI | Nuix Script | FTK |
|---|---|---|---|
| Total unique items | 22(10) | 23(13) | 14(6) |
| Emails | 13(6) | 15(7) | 14(6) |
| Attachments | 9(4) | 13(4) | 0 |
| Desktop files | 0 | 4 | 0 |

Removing the overlapping items between the Nuix GUI and the Nuix Script presented a difference of 10 items. The Nuix script could find these extra 10 items with the use of applying file relation searches. An analysis on these items showed that:

- E-mail 8, the forwarded e-mail to the friend, which originally came from the lawyer was now recognised.
- The result included 4 files from the user's desktop folders, except for File3 on the desktop on the Friend computer. This is a result of the fact that the script does not search for related files that are send to the lawyer.
- File 2 is now included which was send in E-mail 9 to the Friend, which originally was flagged privileged because it was send from the Lawyer to the Suspect in e-mail 3.

Based on the comparison between the Nuix GUI and Nuix script, the addition of the related item search based on shingles improved the privilege data search by 31.25% in this test case.

The experimental results showed that the Nuix script, which was started with a single e-mail address as parameter, could thoroughly filter the privileged information. In comparison to the Nuix GUI and FTK, programmatically removing privileged items resulted in a minimum of 30% better results. This test indicates that extending a standard search to include file content relations will most likely increase the depth of the search in the digital evidence.

The Nuix script 'shields' the person who is working on the privileged information from reading any content. When the Nuix script is started on a file or e-mail, all related files are automatically processed, blocked and exported. In comparison with the traditional way, when the filtering is done manually, the person may still be presented with the list of files or e-mails by the forensic tool after the filter is made. This list will most likely contain the subject of the e-mail or document. The script solves this problem by not requiring any interaction after the start.

The limitation of FTK to filter on the attachments of e-mail communication resulted in a low 50% lower filter result for FTK in comparison to Nuix. This result is unexpected, as FTK is currently the main tool used for digital forensics and the filtering of privileged files.

Out of the two tools investigated, only the commercially tools FTK and Nuix implemented ways to block items from being displayed in the interface. Both implementations are still inadequate, as privileged items should be fully blocked in a case for all users and it should be made difficult to re-add or view these files.

The Nuix script can filter out privileged items as quickly as the software and hardware on which Nuix runs will let it. All steps are automated and require no human interaction after setting the first parameters. In large cases this could decrease the time required for filtering privileged items.

## VI. CONCLUSION AND FUTURE WORK

The results of this paper shows that digital forensic tools are built for extracting as much information as possible, and do a good job at it. But functionally they are still limited when it comes to blocking certain information that is legally not allowed to be viewed. As this is a requirement by law in many countries, it can be questioned if privileged information does not often occur in criminal investigations, it is not recognised or perhaps ignored, or if there are other reasons why forensic tools have implemented this in such a limited way. The script developed for Nuix as part of this paper showed in a test case an increase of 30% when privileged items are compared and related based on the content. This content based relation functionality is currently limited to the forensic tool Nuix, as no other tested forensic tools had the possibility to make such comparisons. This reduces the amount of privileged files law enforcement might still encounter after a case has been filtered for privileged data. We are also looking at other approaches for reducing the amount of files as mentioned in [24][25]. We are also extending our experiments on mobile devices especially focusing on digital forensic tools for VoIP acquisition [26] as well as examining privileged data in GIS application [27] and in privacy web browser [28].

The developed Nuix script in its current form can be easily added to Nuix and is a functional extension to the forensic tool fto deal with privileged information inside a Nuix case. Future feedback from other investigations and the different roles involved in the filtering process, from investigators to the magistrate, will be required to mature and further test the technical and legal requirements.